\documentclass[showpacs,preprintnumbers,amsmath,amssymb]{revtex4} 
\usepackage{graphicx}
\usepackage{dcolumn}
\usepackage{bm}
\usepackage{natbib}
\usepackage{amssymb,amsmath}
\usepackage{longtable}
\usepackage{txfonts}
\usepackage{romannum}
\usepackage{floatrow}
\usepackage{subfigure}
\usepackage{lipsum}
\usepackage[driverfallback=dvipdfm]{hyperref}

\hypersetup{pdftitle = The title of my PDF, pdfauthor = My name, pdfsubject= The subject, pdfkeywords = keyword1 keyword2 keyword3}
\hypersetup{colorlinks = true, linkcolor = green, anchorcolor = red, citecolor = blue, filecolor = red, pagecolor = red, urlcolor = red}

\begin{document}

\title{A Study of the Abundance of Low-Z Elements in\\ the Sun During its Whole Predicted Life}
\author{\firstname{M.\,H.}~\surname{Talafha}}
\email{m.talafha@astro.elte.hu}
\affiliation{Department of Astronomy, Faculty of Science, E$\ddot{o}$tv$\ddot{o}$s Lor$\acute{a}$nd University, Budapest, 1117, Hungary.}
\author{\firstname{M. A.}~\surname{Al-Wardat}}
\affiliation{Department of Physics, Faculty of Science, Al al-Bayt University, Mafraq, 25113, Jordan.}
\author{\firstname{N. M.}~\surname{ Ershidat}}
\affiliation{Department of Physics, Faculty of Science, The University of Jordan, Amman, 11942, Jordan.}


\date{\today}

\begin{abstract}
The study of the elemental composition of stars and galaxies is a key topic for understanding their origin and evolution. 
In this study, we present the results of the calculation of solar abundances of the isotopes $^{1}$H, $^{4}$He, $^{12}$C, $^{14}$N, $^{15}$O, $^{16}$O, $^{17}$O, and $^{18}$O during the four phases of the solar life; Hydrogen burning, Onset of rapid growth and red giant, Helium burning and Helium exhaustion. The open source package nucnet-tools from the Webnucleo Group in Clemson University, SC, USA was used for this purpose. The results for all isotopes are listed in tables for future use. Abundances found, globally, agree fairly well with those predicted in the literature. Results obtained for the last two phases have no equivalent elsewhere.
\end{abstract}


\maketitle
\section{Introduction}

The theory of stellar evolution is by now well established, after more than sixty years of continuous development, and its main predictions confirmed by various empirical tests \citep{2006essp.book.....S}. Thanks to the works of Bethe \citep{1939PhRv...55..103B},   \citep{1957RvMP...29..547B}, and many others, the study of the elemental composition of stars and galaxies, revealed to be a key topic for understanding the origin and evolution of these remote objects \citep{2011eab..book.....G},\citep{1997RvMP...69..995W}. Tools for understanding the stellar evolution are 1) atomic spectroscopy, which allows the determination of the elemental abundances in stars, interstellar gas, and galaxies, and 2) nuclear mass spectroscopy, which is used for precise determination of most isotope abundances in Earth, lunar, and meteoritic samples.
The \textit{abundance} of a given element is the relative amount (or fraction) of this element in a sample of matter. It can be expressed either:
- in relative terms with respect to the amount of some key element, e.g., the most abundant one, in the sample. 
- in absolute terms (definition adopted in this work), i.e., with respect to the total amount of matter in the sample. The abundance is expressed in terms of mass fractions; i.e. the mass fraction of a given nuclide to the total mass of the star.
Measurements of oxygen abundances in stars or in presolar grains of different ages provide important clues about the chemical evolution of the Galaxy \citep{2008RvMG...68...31M}. The isotopes of oxygen are a crucial diagnostic tool of nucleosynthesis and Galactic chemical evolution. This is due in large part to oxygen’s high abundance. $^{16}$O is the third most abundant element in the  solar system, though considerably less abundant than hydrogen and helium (principally produced in primordial nucleosynthesis). According to the compilation of Lodders \citep{2003ApJ...591.1220L} oxygen’s abundance is nearly equal to that of all the other heavy elements (elements with atomic number greater than that of helium) combined \citep{2007prpl.conf..849Y}.

Despite the low relative abundances of $^{17}$O and $^{18}$O, they are still sufficiently abundant that they can be measured in stellar atmospheres. Oxygen isotopic abundances are also readily measured in presolar grains in meteorites and interplanetary dust particles. These measurements provide key constraints on nucleosynthesis and stellar evolution \citep{2009LanB...4B...44L}.
The fact that its three stable isotopes are predominantly made in different burning epochs during the life of a star, contributes to oxygen’s importance as a diagnostic tool. $^{17}$O is predominantly made in hydrogen burning, $^{18}$O is predominantly made in the early stages of helium burning and $^{16}$O is made in the later part of helium burning. Thus, the isotopic composition of a stellar atmosphere or a presolar grain provides clues to the stellar region in which the major part of the material was synthesized \citep{2004ApJ...611..452E}.

Finally, a third significant aspect of oxygen’s role, as a diagnostic tool of nucleosynthesis and galactic chemical evolution, is the primary versus secondary nature of the different isotopes. $^{16}$O is a primary isotope, that is, an isotope that can be synthesized in a star initially composed only of hydrogen, while $^{17}$O and $^{18}$O are secondary isotopes, which means that their formation requires pre-existing seed nuclei from previous stellar generations \citep{2008RvMG...68...31M}.
As a consequence, the abundance of $^{16}$O relative to those of $^{17}$O and $^{18}$O changes with time in the Galaxy’s history. 
4.56 billion years ago, our sun started to burn hydrogen and entered the main sequence phase. It will continue in this phase for at least 6.34 billion years until the hydrogen runs out {http://www.astronomy.ohio-state.edu/~pogge/}. In this paper we present the theoretical calculations of the abundances of the following isotopes: $^{1}$H, $^{4}$He, $^{12}$C, $^{14}$N, $^{15}$O, $^{16}$O, $^{17}$O, and $^{18}$O; over the "whole lifetime" of the sun, i.e. the four phases: Hydrogen burning, Old age, Rapid Growth and Red Giant, Helium burning, and Helium Exhaustion {http://www.astronomy.ohio-state.edu/~pogge/}. To our knowledge, except \citep{r:Talafha}.
\section{Methodology}
\subsection[Calculating the abundance]{Calculating the abundance}
In order to calculate the abundance of a given element; one, first, has to define the series of reactions (network) leading to its production and destruction. This leads to a system of coupled differential equations, known as Bateman equations \citep{1989PhT....42a..78K}. These equations are, generally, solved numerically using specific temperature and density as initial conditions. The open-source package nucnet-tools, initiated and developed by the Webnucleo project headed by Bradley S. Meyer, Astronomy and Astrophysics group at Clemson University, South Carolina, USA {https://sourceforge.net/p/libnucnet/home/Home/}, was used for this purpose. In addition, nucnet-tools allows the determination of the abundance of elements in stars. The sun is being taken as an example in many of the scripts composing the package \citep{2012nuco.confE..96M}. In this work, calculation of the abundance of low-Z elements, namely $^{1}$H, $^{4}$He, $^{12}$C, $^{14}$N and the oxygen isotopes $^{15}$O, $^{16}$O, $^{17}$O, and  $^{18}$O, in the core of the sun is achieved using "nucnet tools". A series of scripts is used in order to calculate the network leading to the production and destruction of an isotope. The Newton-Raphson Method is used for solving the system of equations required for the calculation of the abundance of the isotope, in an iterated method until a defined convergence is reached. The procedure is detailed and well explained in \citep{2012nuco.confE..96M}.\\
As mentioned above, the temperature and density are the main parameters for initializing the calculation. They both, mainly, depend on the expansion timescale ($\tau$) which can be defined as \citep{2012nuco.confE..96M}:\\
\begin{equation}
\frac{1}{\tau} = - \frac{1}{\rho _\circ} \frac{\partial \rho}{\partial t}
\end{equation}

So that the density changes as:

\begin{equation}
\rho (t) = \rho _\circ exp (-t/ \tau)
\end{equation}
and the temperature T$_{9}$ = T/10$^{9}$ K is such that $\rho\propto T^{3}_{9}$\citep{2012nuco.confE..96M}.\\ 

Here, the expansion time scale,$\tau$, is taken equal to infinity. This choice means that the temperature T and density ($\rho$), in the center of the sun, remain constants with time. The initial values of T and $\rho$ are, respectively, 1.548 $\times$ 10$^{7}$ K, 1.505 $\times$ 10$^{2}$ g/cm$^{3}$ at time t = 10.9 billion years.

Appropriate nuclides and reactions data from the JINA (Joint Institute for Nuclear Astrophysics) database ~\citep{2017nuco.confb0203C} are used when needed.

\subsection{Production and destruction of oxygen isotopes}

The following reactions contribute to the production of oxygen isotopes \citep{2008RvMG...68...31M}: $^{14}_{7}N$(\textit{p},$\gamma$)$^{15}_{8}O$, $^{19}_{9}F$(\textit{p},$\alpha$)$^{16}_{8}O$, $^{15}_{7}N$(\textit{p},$\gamma$)$^{16}_{8}O$, $^{12}_{6}C$($\alpha$,$\gamma$)$^{16}_{8}O$, and the two decay reactions:  $^{17}_{9}F\longrightarrow^{17}_{8}O + e^{+} + v_{e}$ and $^{18}_{9}F\longrightarrow^{18}_{8}O + e^{+} + v_{e}$. \\

The decay reaction, always has constant reaction rates. "nucnet-tools" allows the calculation of such reaction rates.We find, for example, that the reactions $^{17}_{9}F\longrightarrow^{17}_{8}O + e^{+} + v_{e}$ and $^{18}_{9}F\longrightarrow^{18}_{8}O + e^{+} + v_{e}$ have the rates 1.0746$\times$10$^{-2}$ and 1.0518$\times$10$^{-4}$ respectively.

It is worth noting that the reactions rates in the output are the product 
N$_{A}$ $\left<v\right>$ which has units of cm$^{3}$ mole$^{-1}$ s$^{-1}$,where N$_{A}$ is the Avogadro's number,  is the reaction's cross-section and \textit{v} is the velocity of the projectile.
The destruction of oxygen isotopes occurs by several reactions, such as:
$^{16}_{8}O$(\textit{p},$\gamma$)$^{17}_{9}F$, $^{17}_{8}O$(\textit{p},$\gamma$)$^{18}_{9}F$,
$^{17}_{8}O$(\textit{p},$\alpha$)$^{14}_{7}N$, $^{18}_{8}O$(\textit{p},$\gamma$)$^{19}_{9}F$, and $^{15}_{8}O \longrightarrow ^{15}_{7}N + e^{+} + v_{e}$.
Some of these reactions involve a proton or an alpha giving a new heavier element, and some of them are decay reactions which have a constant reaction rate at any temperature such as $^{15}_{8}O \longrightarrow ^{15}_{7}N + e^{+} + v_{e}$ with rate 5.6816$\times$10$^{-3}$.

\section{Results}
\label{Atm}
Calculations were performed over the "whole lifetime" of the sun. The first two columns in Table~\ref{tbl-1} show the consecutive phases of our sun and their lifetimes, as in {http://www.astronomy.ohio-state.edu/~pogge/}. The initial values of the important parameters for achieving the iterations, namely; the temperature (10$^{9}$ K), density (g/cm$^{3}$), and the value of abundance to start with, are also shown in Table ~\ref{tbl-1}.

\textit{Preliminary note:} the figures herein illustrate the most significant results. The initial values for the abundance of the studied isotopes are shown in Table~\ref{tbl-2}. All results are compiled in a single table (Table~\ref{tbl-3}). 

Calculations are run with an infinite expansion timescale ($\tau$). In addition the initial conditions in Table~\ref{tbl-2} are assumed. The electron screening and nuclear statistical corrections are neglected in these calculations. 
The radius of the sun changes during phases, so that, the density changes accordingly. The density in phases II, III, and IV are taken to be $\rho$ = $\rho_{\circ}\times r^{3}$, where r is the radius of the sun (in units of the actual radius, R$_{\circ}$) at the beginning of each phase {http://www.astronomy.ohio-state.edu/~pogge/}

\begin{table*}
\small
\caption{Phases of the Sun. Initial conditions for each phase in the sun's lifetime.\label{tbl-1}}
\begin{tabular}{@{}lccll@{}}
\tableline
Phase & Duration & Temp.(10$^{9}$ K) & Density(g/cm$^{3}$) & Abundance\\
\hline
Hydrogen Burning & 10.9 Gy & 0.0154 & $\rho_{0}$ = 150.5 & Lodders\\
Old age,  & 1301 My & 0.0154 & $\rho$ = $\rho_{0}\times(2.3)^{3}$ = 1831.1335 & last step in H burning\\
Rapid Growth and RG &&&&\\
Helium Burning & 110 My & 0.1 & $\rho$ = $\rho_{0}\times(9.5)^{3}$ = 129034.9375 & last step in previous phase\\
Helium Exhaustion & 20 My & 0.6 & $\rho$ = $\rho_{0}\times(18)^{3}$ =877716 & last step in He burning\\
\hline
 & 12.331 Gy
 \end{tabular}

\end{table*}

\begin{table*}
\small
\caption{Initial values (t = 0) of the abundance of the studied elements from \citep{2003ApJ...591.1220L}} \label{tbl-2}
\begin{tabular}{@{}lllllllll}
\tableline
 & $^{1}$H & $^{4}$He & $^{12}$C & $^{14}$N & $^{15}$O & $^{16}$O & $^{17}$O & $^{18}$O \\
\hline
Initial values (t = 0) & 0.7109 & 0.2741 & 0.0025 & 0.0008 & 1.18$\times 10 ^{-25}$ & 0.0066 & 2.62$\times 10^{-6}$ & 1.49$\times 10 ^{-5}$
\end{tabular}
\end{table*}

\subsection{Phase I: The H Burning Phase}

Hydrogen burning is the first phase in the lifetime of the sun, where hydrogen burns to produce helium, it will continue for 10.9 Gy.
Figures 1 to 5 show, the mass fractions of the isotopes $^{1}$H with $^{4}$He, $^{12}$C, $^{14}$N and the oxygen isotopes $^{16}$O, $^{17}$O and $^{18}$O, respectively, during this phase. 
One can see from these figures, that for the last time step in the calculation (10.9 Gy), the abundance for $^{1}$H is X = 0.1121, for $^{4}$He is Y = 0.8725 and is 6.86$\times$10$^{-6}$ for $^{12}$C. These results together with the abundances for $^{14}$N and the oxygen isotopes ($^{15}$O, $^{16}$O, $^{17}$O and $^{18}$O) are shown in Table~\ref{tbl-3}.

\subsection{Phase II: Old age, Onset rapid growth, and red giant (1.301 Gy)}
After the hydrogen core is exhausted, the helium atoms become unstable and start to collapse under their own weight which makes the core hotter and denser while the remaining hydrogen gets moved out into a thin shell surrounding the helium core. The next 700 My (phase ending at 11.6 Gy) show a slow evolution where the brightness remains constant and the radius gets bigger; the sun becomes a subgiant star.
After that, the subgiant Sun starts a rapid growth in size for 601 My (after 12.201 Gy from its birth). At that time, the stellar wind begins and the outer parts of the sun's envelope get lost to the wind. The red giant phase starts at 12.233 Gy, the sun becomes brighter and bigger to engulf Mercury, the helium core will reach a temperature of 100 Million degrees and helium fusion ignites in the core producing carbon and oxygen {http://www.astronomy.ohio-state.edu/~pogge/}. 
Calculations show no significant variation of the mass fraction of all the studied elements over the total period of 1301 My. Thus the two previous periods were grouped in one phase (phase II of 1.301 Gy).
Calculations show that for the last time step (1301 My) the abundance for $^{1}$H is X = 1.62 $\times$10$^{-13}$, for $^{4}$He is Y = 0.8736. The corresponding abundance of $^{12}$ C is 6.87 $\times$  10$^{-6}$. These results together with the abundances for $^{14}$N and the oxygen isotopes ($^{15}$O, $^{16}$O, $^{17}$O, and $^{18}$O) are shown in Table~\ref{tbl-3}.\\
As expected, the star runs out of hydrogen and the main source of energy is the helium burning.

\subsection{Phase III: Helium Burning Phase}

At 12.201 Gy the red giant settles down for a 110 My of stability as a helium burning star. At the end of this period, the giant's core finally runs out of helium and the C-O ash begins to collapse rapidly {http://www.astronomy.ohio-state.edu/~pogge/}.
Table~\ref{tbl-3} shows the mass fractions of the studied elements during this phase.

\subsection{Phase IV: Helium Exhaustion Phase}

Finally, and the calculations stop here, the same procedure is used for the He exhaustion phase which lasts for 20 My.
The mass fractions of all the studied elements during the helium exhaustion phase are shown and discussed in Table~\ref{tbl-3}.
At the last time step of the helium exhaustion phase the abundance of elements will be X = 8.22782 $\times$10$^{-34}$ for $^{1}$H, Y = 3.469325$\times$10$^{-20}$ for $^{4}$He and 2.20796 $\times$10$^{-14}$ for $^{12}$C. 
Calculations were also done for $^{19}$O, following the same procedure as for the other isotopes. The calculations give infinitesimally small mass fractions for all the four phases.
For simplicity and clarity, the choice was made to add the discussion of this work's results as part of Table~\ref{tbl-3}.

\begin{figure}[ht!]
     \begin{center}
        \subfigure[Abundance of $^{1}$H and $^{4}$He in the H-burning phase]{%
            \label{fig1}
            \includegraphics[width=0.5\textwidth]{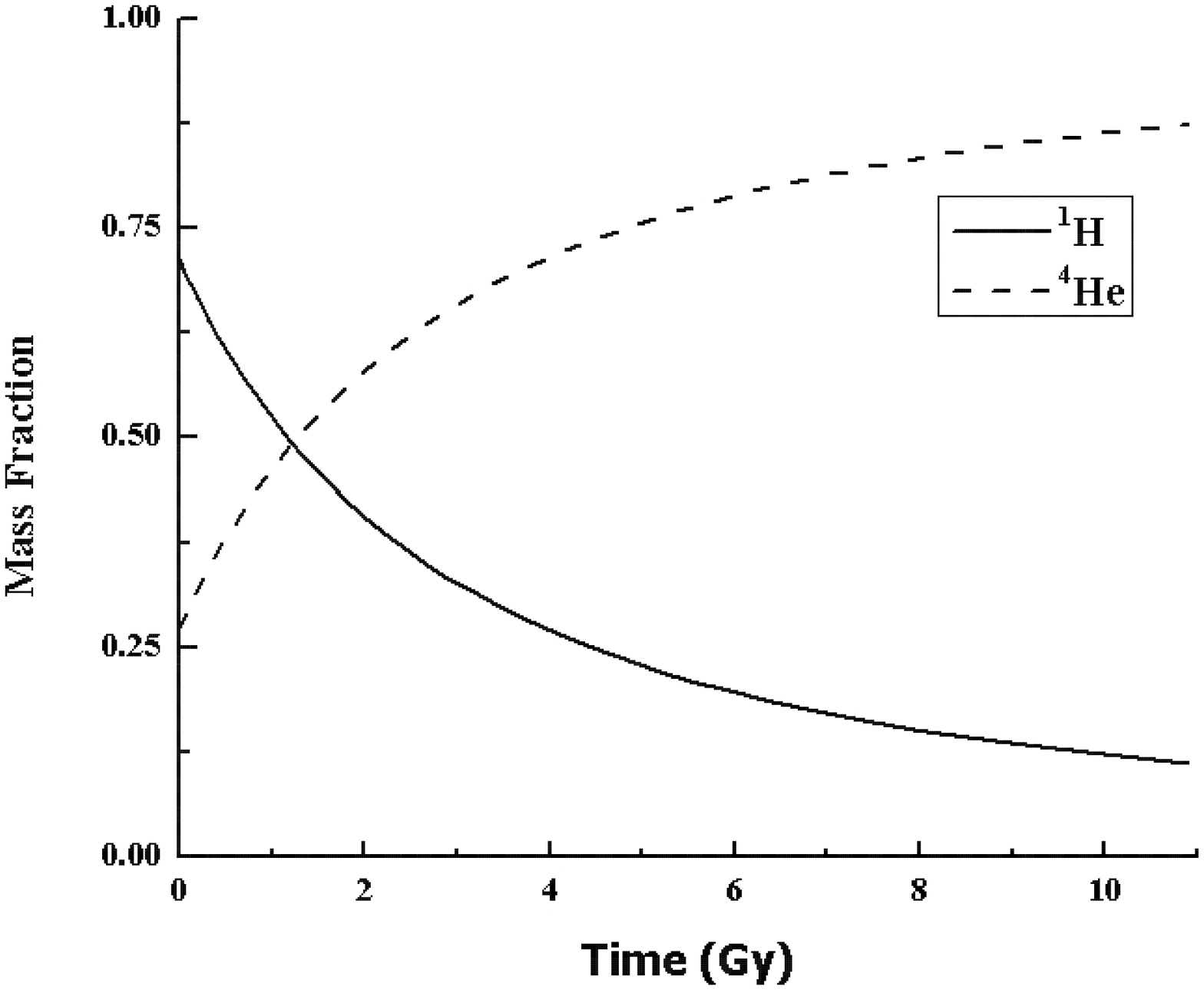}
        }%
        \subfigure[Abundance of $^{12}$C in the H-burning phase]{%
           \label{fig2}
           \includegraphics[width=0.5\textwidth]{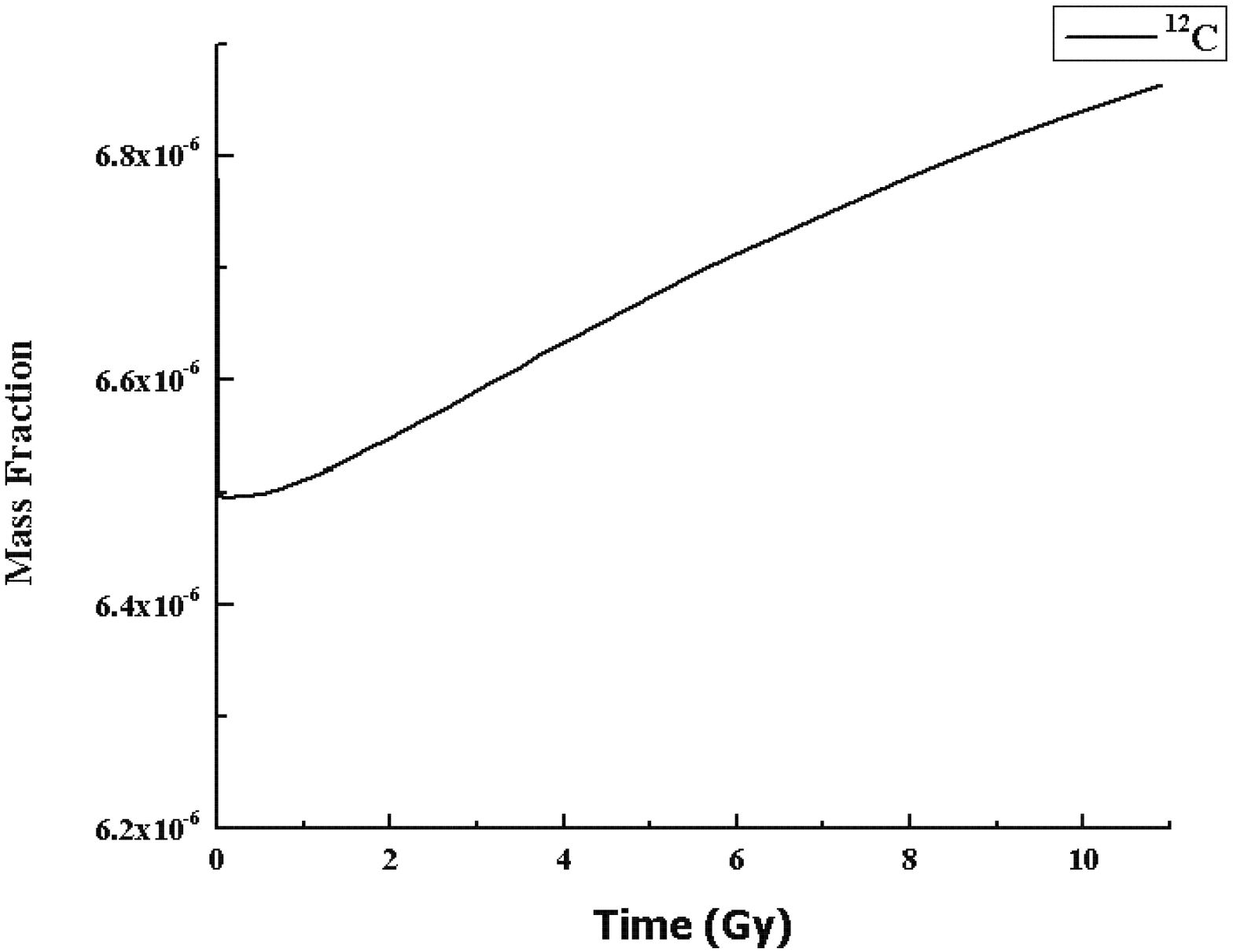}
        }\\ 
        \subfigure[Abundance of $^{14}$N in the H-burning phase]{%
            \label{fig3}
            \includegraphics[width=0.5\textwidth]{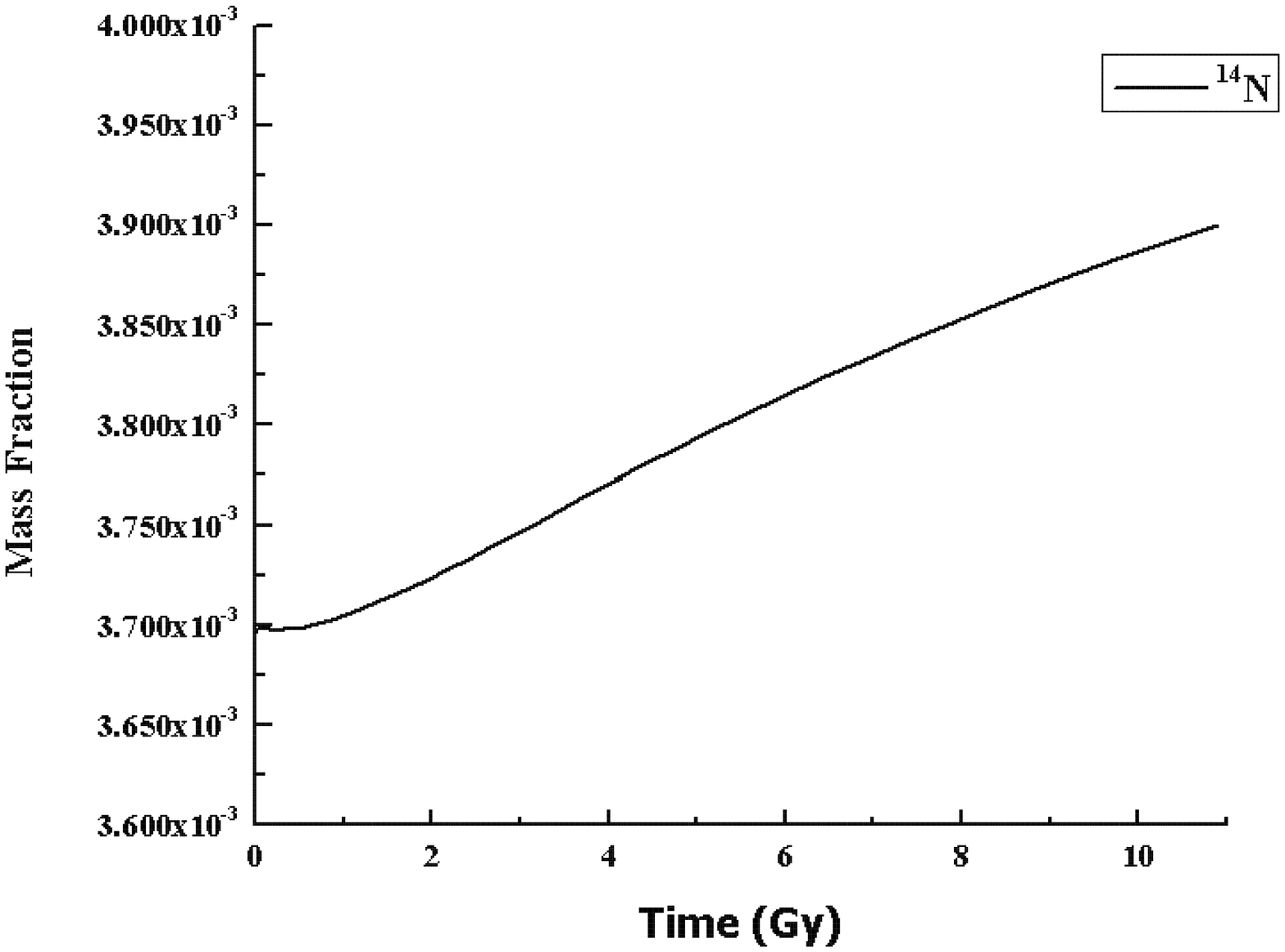}
        }%
        \subfigure[Abundance of $^{16}$O in the H-burning phase]{%
            \label{fig4}
            \includegraphics[width=0.5\textwidth]{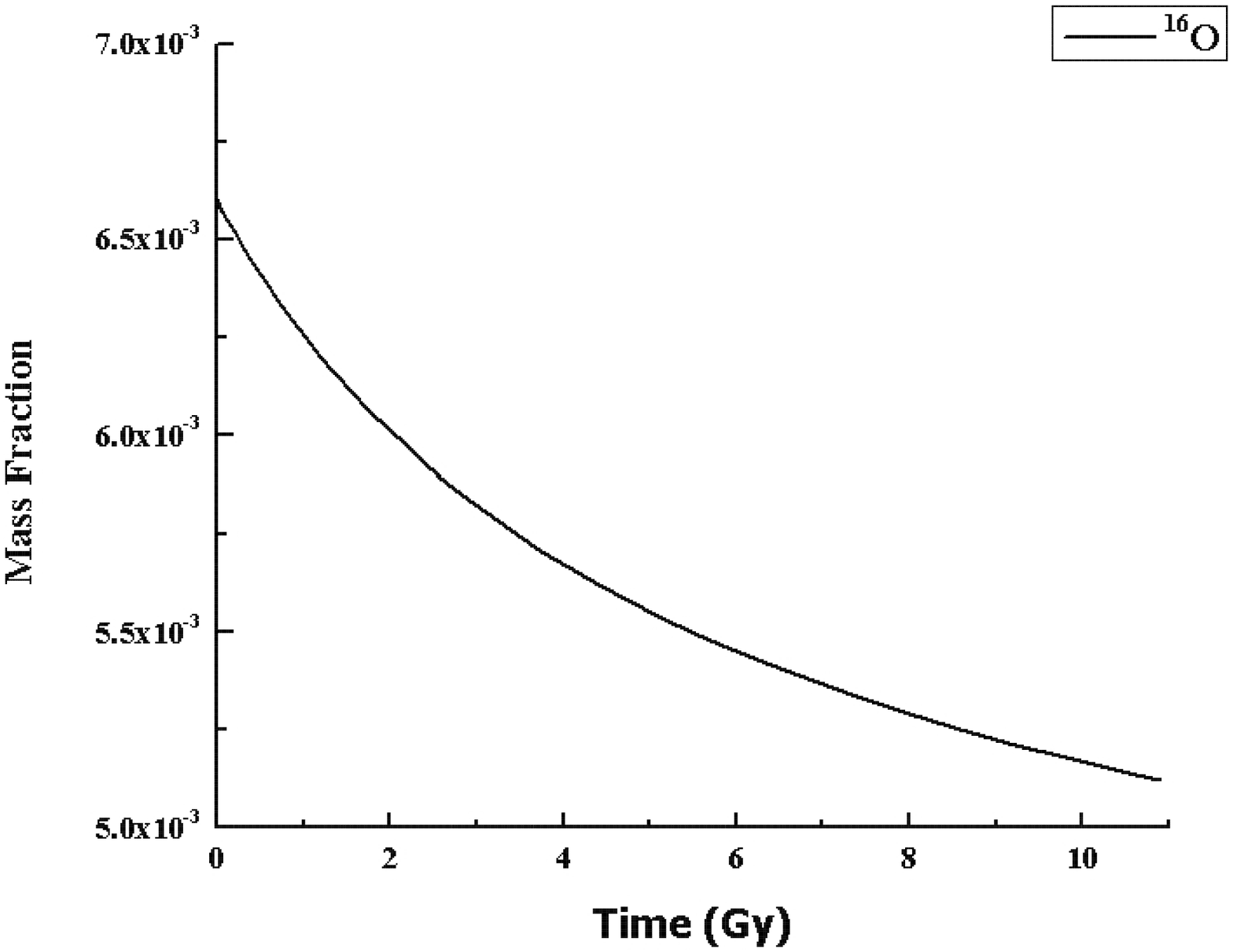}
        }\\%
        \subfigure[Abundance of $^{17}$O and $^{18}$O in the H-burning phase]{%
            \label{fig5}
            \includegraphics[width=0.5\textwidth]{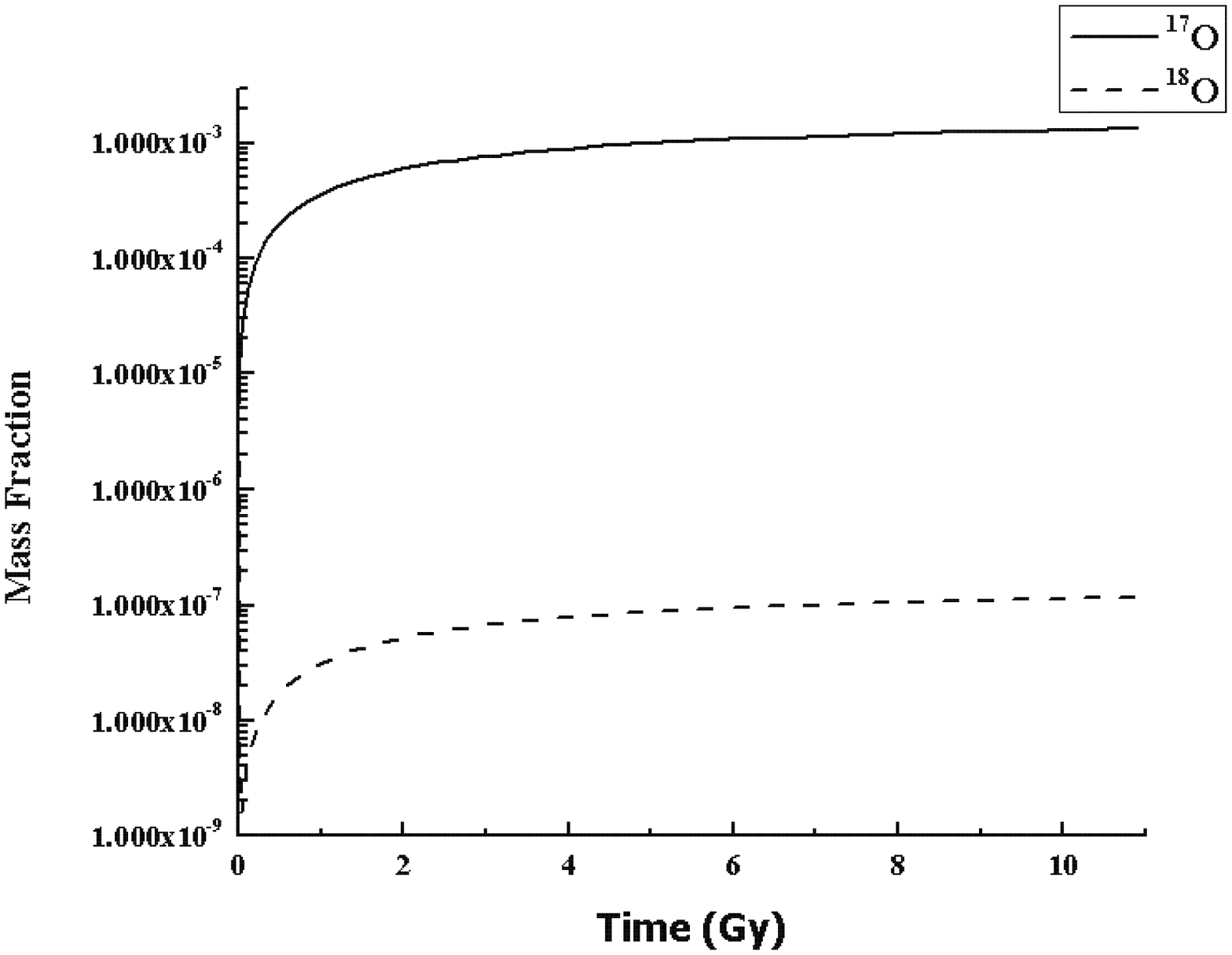}
        }%
    \end{center}
    \caption{%
        Abundances of Low-z elements during the five phases of the sun.
     }%
   \label{fig6}
\end{figure}

\section{Conclusions}

In this work, the abundance of the isotopes $^{1}$H, $^{4}$He, $^{12}$C, $^{14}$N, $^{15}$O, $^{16}$O, $^{17}$O, and $^{18}$O in the core of our sun, were calculated over a period of 12.331 Gy. The main tool used is the open source package nucnet-tools from the Webnucleo Group in Clemson University, South Carolina, USA.
The predicted lifetime of the sun is divided into four main phases; namely 1) Hydrogen burning, 2) Lively old age, Onset rapid growth and red giant, 3) Helium burning and 4) Helium exhaustion (Table 1).
The initial abundances of the elements were taken from Lodders compilation for proto-sun. The obtained results, in this work, confirm the predictions by nuclear physics models.
To avoid redundancy, the reader is referred to the last column in Table~\ref{tbl-3} which resumes all the conclusions that can be driven from this work. But globally, one can say that the calculations are consistent with the theoretical predictions of the abundance of studied isotopes, and especially the oxygen isotopes, in the core of the sun.

\newpage
\begin{table*}{}
\caption{Abundance of the studied elements at the end of each of the 4 phases and major observations}
\label{tbl-3}
\begin{tabular}{@{}lllllllll}


Element & Phase & Abundance &  \\
 & I & 0.1121 &  Mass fraction decreases in phases I, II. At the end of phase IV \\
  $^{1}H$      & II & 0.0441804 &  there is almost no hydrogen. \\
        & III   & 1.23$\times 10 ^{-25}$ &    \\
        & IV    & 1.99$\times 10 ^{-26}$ &     \\
\hline
 & I     & 0.8725      & The mass fraction of 4He increases and  reaches a maximum at    \\
 $^{4}He$ & II    & 0.940427   & the end of phase III. 20 My after that, i.e. at the end of the   \\
        & III   & 0.3187 & last phase (IV) starts,the so-called, Helium flash. This phase \\
        & IV    & 8.78$\times 10 ^{-18}$ & is beyond our study.    \\
\hline
 & I     &  6.86$\times 10 ^{-6}$  & The mass fraction of $^{12}$C increases very slowly and reaches    \\
$^{12}C$  & II    & 6.9$\times 10 ^{-6}$   & a maximum at the end of phase III. In the last phase (IV) \\
        & III   & 0.3893  & $^{12}$C  is still present in a fraction much larger than that of $^{1}$H.   \\
        & IV    & 0.00198 &     \\
\hline
 & I     & 0.0039 & The mass fraction of $^{14}$N is almost stable over the first    \\
$^{14}N$  & II    & 0.004    & two phases (12.201 Gy) then it increases and reaches     \\
        & III   & 8.37$\times 10 ^{-3}$ & a maximum at the end of phase III. In the last phase    \\
        & IV    & 5.49$\times 10 ^{-11}$ & (IV) $^{14}$N almost vanishes.    \\
 \hline
 & I     & 1.60$\times 10 ^{-17}$      & The mass fraction of $^{15}$O is almost stable over the four phases.    \\
 $^{15}O$  & II    & 7.10$\times 10 ^{-17}$  & It constantly decreases to a very small fraction. This is expected   \\
        & III   & 1.72$\times 10 ^{-26}$ & since this isotope is unstable (half-life = 122 s)   \\
        & IV    & Less than 1$\times 10 ^{-30}$ &     \\
\hline
 & I     & 5.10$\times 10 ^{-3}$    & The mass fraction of $^{16}$O is almost stable till the end of phase III.  \\
$^{16}O$  & II    & 4.81$\times 10 ^{-3}$   & In the last 20 My, a drastic decrease is found. Nevertheless,  \\
        & III   & 0.2775 &  this isotope is still present at the end of the four phases.    \\
        & IV    & 0.030 &     \\

\hline
 & I     & 0.0013    & The mass fraction of 17O is almost stable till the end of phase II.    \\
$^{17}O$  & II    & 0.0015   &  This stable isotope is still present at the end of the first three    \\
        & III   & 1.08$\times 10 ^{-9}$  &  phases. Nevertheless a drastic decrease is found in the last phase,  \\
        & IV    & 1.20$\times 10 ^{-9}$ &   as expected.     \\

\hline
 & I     & 1.17$\times 10 ^{-7}$    & The mass fraction of $^{17}$O is very small compared to $^{16}$O.  Its    \\
$^{18}O$  & II    & 1.33$\times 10 ^{-7}$   & abundance is larger in the first two phases, as mentioned in    \\
        & III   & 7.80$\times 10 ^{-4}$ & the text. The obvious variations from phase to another is due   \\
        & IV    & 2.55$\times 10 ^{-11}$ & to the reactions of its production and destruction. A clarification  \\
        & & &  can be obtained thanks to a future study on the calculation of \\
        & & & reaction rates of the oxygen isotopes \\
\hline       
$^{19}O$  & all phases    & Infinitesimal traces  &     \\


\end{tabular}
\end{table*}

\newpage

\newpage

\bibliographystyle{mn2e}
\bibliography{AbundanceSun}
\end{document}